%% file: main.tex
\documentclass[%
reprint,
superscriptaddress,
nofootinbib,
 amsmath,amssymb,
 aps,
]{revtex4-2}

\usepackage[english]{babel}
\usepackage{xcolor}
\usepackage[linktocpage, colorlinks, citecolor=blue, linkcolor=blue, urlcolor=blue]{hyperref} 
\usepackage{enumerate}  
\usepackage{float}
\usepackage{calc}
\usepackage{dutchcal}
\usepackage{revsymb}
\usepackage{comment}
\usepackage{upgreek}
\usepackage{IEEEtrantools}
\usepackage{MyCommands}
\usepackage[normalem]{ulem}

\usepackage{amsfonts, amssymb, amsthm, mathtools, mathrsfs}
\usepackage{physics}

\usepackage{graphicx}
\usepackage{dcolumn}
\usepackage{bm}

\newcommand{\orcid}[1]{\href{https://orcid.org/#1}{\includegraphics[width=8pt]{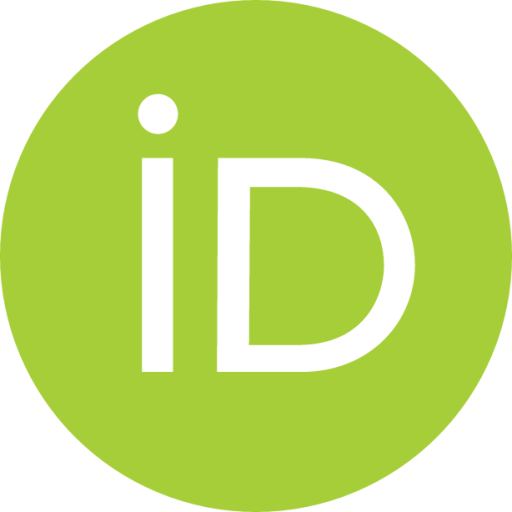}}}

\begin{document}

\title{Unifying Ordinary and Null Memory}

\author{Lavinia Heisenberg}
\email{l.heisenberg@thphys.uni-heidelberg.de}
\affiliation{Institut f\"{u}r Theoretische Physik, Philosophenweg 16, 69120 Heidelberg, Germany}
\author{Guangzi Xu}
\email{xuguan@student.ethz.ch}
\affiliation{Institute for Theoretical Physics, ETH Z\"{u}rich, Wolfgang-Pauli-Strasse 27, 8093, Z\"{u}rich, Switzerland}
\author{Jann Zosso\orcid{0000-0002-2671-7531}}
\email{jann.zosso@nbi.ku.dk}
\affiliation{Institute for Theoretical Physics, ETH Z\"{u}rich, Wolfgang-Pauli-Strasse 27, 8093, Z\"{u}rich, Switzerland}
\date{\today}

\begin{abstract}
Based on a recently proposed reinterpretation of gravitational wave memory that builds up on the definition of gravitational waves pioneered by Isaacson, we provide a unifying framework to derive both ordinary and null memory from a single well-defined equation at leading order in the asymptotic expansion. This allows us to formulate a memory equation that is valid for any unbound asymptotic energy-flux that preserves local Lorentz invariance. Using Horndeski gravity as a concrete example metric theory with an additional potentially massive scalar degree of freedom in the gravitational sector, the general memory formula is put into practice by presenting the first account of the memory correction sourced by the emission of massive field waves. Throughout the work, physical degrees of freedom are identified by constructing manifestly gauge invariant perturbation variables within an SVT decomposition on top of the asymptotic Minkowski background, which will in particular prove useful in future studies of gravitational wave memory within vector tensor theories.

\end{abstract}

\maketitle


\input{MainBody}

\newpage
\bibliographystyle{utcaps}
\bibliography{references}

\clearpage

\end{document}

%% file: MainBody.tex
\section{\label{sec:Intro}Introduction}

The revolution in gravity physics at the beginning of the last century was marked by a tremendous advance in the conception of the notions of space and time. Using worldlines of free test particles as basic probes of spacetime, the statement of the equivalence principle that fundamentally there exist no test particles free of the gravitational interactions entails the far-reaching consequence of viewing spacetime as a curved and dynamical concept. More precisely, inertial observers of special relativity tracing out the flat and rigid fabric of Minkowski spacetime are to be replaced by the notion of freely falling frames, only locally recovering the Minkowski structure. Famously, Einstein's theory of general relativity (GR) emerged as the preeminent framework for understanding gravity.

Yet, despite the remarkable success of GR, there remain several open questions, in particular in the context of cosmology and the unification with quantum physics, that drive the exploration of the theory space beyond GR. Interestingly, the Einstein equivalence principle together with locality, covariance, and a restriction to four space-time dimensions leaves room for a broader theoretical framework beyond GR that might be worth exploring in this context. While GR is identified through the Lovelock theorem \cite{Lovelock1969ArRMA} as the unique local and covariant low energy gravity theory constructed out of a single gravitational metric on a manifold, the larger theory space of \textit{metric theories of gravity} \cite{Dicke:1964pna,poisson2014gravity,papantonopoulos2014modifications,Will:2014kxa,Heisenberg:2018vsk,Will:2018bme,YunesColemanMiller:2021lky,Heisenberg:2023prj} is characterized through the possibility of describing additional degrees of freedom in the gravitational sector. In doing so, it is however of imminent importance to maintaining the requirement of the existence of a single physical metric that minimally and universally couples to matter, in order to abide to the Einstein equivalence principle.

One popular alternative gravity theory is given by the Horndeski action \cite{Horndeski:1974wa,Nicolis:2008in,Deffayet:2009wt,Deffayet:2009mn,Heisenberg:2018vsk,Kobayashi:2019hrl}, representing the most general metric theory incorporating an extra scalar field in the gravitational sector, with the additional property of exhibiting equations of motion at most at second-order in derivatives per field. This supplementary requirement ensures the absence of any Ostrogradski instability \cite{Ostrogradsky:1850fid,Woodard:2015zca}. 
While such an extra scalar gravitational degree of freedom naturally finds its application in cosmology, its traces should equally be searched for within toady's and tomorrow's gravitational wave data. 

Indeed, the groundbreaking first direct discovery of a gravitational wave signal \cite{LIGOScientific:2016aoc,LIGOScientific:2016lio} opened a new window for testing the nature of the gravitational interaction in entirely new regimes. In this context, the framework of metric theories provides a viable interpretation of the measurements on the basis of a well-defined geodesic deviation equation governed by the electric part of the Riemann tensor associated to the physical metric. In particular, metric theories of gravity therefore also lay out the theoretical basis for discussing the gravitational wave memory effect beyond general relativity \cite{Heisenberg:2023prj}.

Within GR, the memory effect, defined as a permanent geodesic displacement after a gravitational wave burst, was first discovered within the linearized theory, arising due to the presence of unbound matter components within an otherwise isolated system that is emitting gravitational waves \cite{Zeldovich:1974gvh,Turner:1977gvh,Braginsky:1985vlg, Braginsky:1987gvh}. It was subsequently understood that the gravitational waves themselves represent an unbound energy component that results in a memory correction when considering the full non-linear theory in asymptotically flat spacetime \cite{Christodoulou:1991cr,Ludvigsen:1989cr,Blanchet:1992br,FrauendienerJ,Thorne:1992sdb,PhysRevD.44.R2945} (see also \cite{Favata:2008yd,Favata:2008ti,Favata:2009ii,Favata:2010zu,Bieri:2013ada,Ashtekar:2014zsa,Strominger:2017zoo,Compere:2019sm,Garfinkle:2022dnm,DAmbrosio:2022clk}). With a precise notion of asymptotic null infinity at hand, it has proven useful to let go of the historic classification between ``non-linear'' and ``linear'' memory, in favor of a distinction between \textit{null} and \textit{ordinary} memory, denoting memory components that are sourced by unbound energy-momentum fluxes that do, respectively don't reach the null boundary at asymptotic infinity \cite{Bieri:2013ada}. More precisely, the Bondi-Metzner-Sachs (BMS) flux-balance laws \cite{Bondi:1962px,Sachs:1962wk,Geroch:1977jn,Ashtekar:1981bq,Heisenberg:2023mdz} naturally distinguish between massless memory source-fluxes traveling at the speed of light and massive unbound components of the system \cite{Christodoulou:1991cr,FrauendienerJ,Ashtekar:2014zsa,Strominger:2014pwa,Tolish:2014bka,Tolish:2014oda,Strominger:2017zoo,Compere:2019gft,DAmbrosio:2022clk,Heisenberg:2023mdz}.

Excitingly, the first detection of the gravitational wave memory effect is soon to be expected \cite{Favata:2009ii,Johnson:2018xly,Yang:2018ceq,Islo:2019qht,Hubner:2019sly,Burko:2020gse,Ebersold:2020zah,Hubner:2021amk,Islam:2021old,Sun:2022pvh,LISA:2022kgy,Grant:2022bla,Gasparotto:2023fcg,Ghosh:2023rbe,Goncharov:2023woe}, at the very least with next-generation ground based detectors such as the Einstein Telescope \cite{Punturo:2010zz,Maggiore:2019uih} or Cosmic Explorer \cite{Reitze:2019iox,Evans:2021gyd}, as well as with the Laser Interferometer Space Antenna (LISA) mission \cite{LISA}. This in particular also motivates the search for a deeper understanding of memory beyond GR to harvest the full potential of gravitational memory in probing the theory of gravity and its implications on the fabric of spacetime. 

Here, we will build up on the novel perspective on displacement memory provided in \cite{Heisenberg:2023prj}, which identified the Isaacson approach to gravitational waves \cite{Isaacson_PhysRev.166.1263,Isaacson_PhysRev.166.1272} as the ideal framework to study gravitational wave memory within generic metric theories. While \cite{Heisenberg:2023prj} primarily focused on null memory by only considering massless field perturbations, the present work will provide a generalization to massive degrees of freedom in the gravitational sector in the particular case of Horndeski gravity. This will not only provide the first description of displacement memory that is sourced by the emission of a massive scalar wave, but also demonstrate a unified derivation of ordinary and null memory within the Isaacson picture. 

After introducing the Horndeski metric theory, Sec.~\ref{sec:MemoryUnified} will review the leading order Isaacson equations of motion, with an emphasis on the importance of a well-defined perturbation theory setup that might have been obscured in the original work. Furthermore, we choose to resolve the associated gauge ambiguity by explicitly identifying the gauge invariant variables within a scalar-vector-tensor (SVT) decomposition of the field perturbations within an assumed asymptotically flat region that preserves Lorentz invariance. Subsequently, Sec.~\ref{Sec:Polarizations} introduces the geodesic deviation equation in generic metric theories of gravity that is both at the root of an unambiguous gravitational wave response, as well as the definition of gravitational wave memory. Moreover, we seize the opportunity to offer a careful discussion of gravitational wave polarizations of Horndeski theory within the SVT decomposition that to our knowledge is hard to come by in the literature. Finally, Sec.~\ref{sec:MemoryHorndeski} will dig into Isaacson's low-frequency back-reaction equation, providing a unified treatment of null and ordinary memory, including a description of gravitational wave memory originating from the emission of a massive Horndeski scalar degree of freedom in the gravitational sector. 

In this work, we choose our metric signature to be $(-,+,+,+)$. Moreover, we will use natural units in which $c=1$.

\section{\label{sec:MemoryUnified}Gravitational Waves in Horndeski Theories}

\subsection{\label{Ssec:Horndeski}Basics of Horndeski Gravity}

Horndeski theory represents the most general scalar-tensor metric theory of gravity with equations of motion that include only up to two derivative operators per field, constructed out of a metric tensor $g_{\mu\nu}$ and a real-valued scalar field $\phi$ \cite{Horndeski:1974wa,Nicolis:2008in,Deffayet:2009wt,Deffayet:2009mn,Heisenberg:2018vsk,Kobayashi:2019hrl}. The action of Horndeski gravity can be written in the following form \cite{Kobayashi:2019hrl}
\begin{eqnarray}
    S = \frac{1}{2\kappa}\int d^4x \sqrt{-g} \left ( \sum_{i=2}^{5} \mathcal{L}_i[g,\phi]\right ) +S_\text{m}[g,\Psi_\text{m}]\,,
    \label{eqn:genprocaaction}
\end{eqnarray}
with $\kappa\equiv 8\pi G$ and where the Lagrangian's $\mathcal{L}_i$ read
\begin{align}
    \mathcal{L}_2  =& \text{  } G_2(\phi, X)  \,, \nonumber \\
    \mathcal{L}_3  =& \text{  } -G_3(\phi, X) \Box \phi \,, \nonumber \\
    \mathcal{L}_4  =& \text{  } G_4(\phi, X)R + G_{4,X}\left [ (\Box \phi)^2 - \nabla_\rho \nabla_\sigma \phi \nabla^\rho \nabla^\sigma \phi \right ]  \,, \nonumber \\
    \mathcal{L}_5  =& \text{  } G_5(\phi, X)G_{\mu\nu} \nabla^\mu \nabla^\nu \phi \nonumber \\
    & - \frac{1}{6}G_{5,X} \left [ (\Box \phi)^3 
    + 2 (\nabla_\rho \nabla^\sigma \phi) (\nabla_\gamma \nabla^\rho \phi) (\nabla_\sigma \nabla^\gamma \phi)  \right. \nonumber \\
    & \left. - 3(\Box \phi) (\nabla_\rho \nabla_\sigma \phi) (\nabla^\sigma \nabla^\rho \phi) \right ]  \,.
\end{align}
The functions $G_i$'s are arbitrary scalar functions depending on the scalars $\phi$ and $X \equiv -\nabla_\mu \phi \nabla^\mu \phi$, and $G_{i,Z} \equiv \partial G_i / \partial Z$ denotes partial derivatives of these scalar functions. Moreover, we further define $\Box \phi \equiv \nabla_\mu \nabla^\mu \phi$ and $S_\text{m}[g,\Psi_\text{m}]$ denotes the matter action in terms of generic matter fields $\Psi_\text{m}$ that are minimally coupled to the physical metric $g$ only. 

As already mentioned, this last requirement is crucial to ensure the metric theory character of Horndeski gravity that is for instance also crucial for the existence of a locally conserved matter energy-momentum tensor
\begin{equation}
    T_{\mu\nu}[g,\Psi_\text{m}]\equiv \frac{-2}{\sqrt{-g}}\frac{\delta S_\text{m}[g,\Psi_\text{m}]}{\delta g^{\mu\nu}}\,.
\end{equation}
Indeed, minimal and universal coupling together with the diffeomorphism invariance of the action ensures its covariant conservation 
\begin{equation}
    \nabla_\mu T^{\mu\nu}=0\,.
\end{equation}
This matter energy-momentum tensor in particular represents a source term for the fields in the gravitational sector within the corresponding equations of motion. For notational clarity, we will denote the metric and scalar field equations as
\begin{subequations}\label{eq:FE}
\begin{align}
    \mathcal{G}_{\mu\nu}[g,\phi]&=\kappa\, T_{\mu\nu}[g,\Psi_m]\,,\label{eq:MetricFE}\\
    \mathcal{J}[g,\phi]&=0\,.\label{eq:ScalarFE}
\end{align}
\end{subequations}
The explicit form of the equations of motion can for instance be found in the Appendix of \cite{Kobayashi:2011nu}. 


As the most general scalar-tensor theory with second-order equations of motion, Horndeski theory encompasses many specific gravity theories beyond GR. Classic examples are Brans-Dicke (BD) theory \cite{Brans:1961sx,Dicke:1961gz,Weinberg1972,poisson2014gravity} that is recovered through the specific choice of functionals
\begin{equation}
    G_2 = \frac{\omega X}{\phi}\,,\;\;G_4 = \phi\,,\;\;G_3 = G_5 = 0\,, \label{eq:BD}
\end{equation}
as well as $f(R)$ gravity \cite{Bergmann:1968aj,Ruzma:1969JETP,Buchdahl:1970MN,Sotiriou:2008rp,DeFelice:2010aj} that is equivalent to choosing
\begin{equation}
    G_2 = f(\phi)-\phi f'(\phi)\,,\;\;G_4 = f'(\phi)\,,\;\;G_3 = G_5 = 0\,, \label{eq:f(R)}
\end{equation}
and scalar Gauss-Bonnet (sGB) gravity \cite{Zwiebach:1985uq,Gross:1986iv} arising from the combination \cite{Kobayashi:2011nu,Kobayashi:2019hrl}
\begin{align}
    G_2 &= X +8f^{(4)}(\phi)X^2(3-\ln X) ,  \nonumber \\
    G_3 &= 4f^{(3)}(\phi)X(7-3\ln X) , \nonumber \\
    G_4 &= 1 +4f^{(2)}(\phi)X(2-\ln X) ,  \nonumber \\
    G_5 &= -f^{(1)}(\phi)\ln X ,  \label{eq:sGB}
\end{align}
where $f^{(n)} \equiv \partial^n f/ \partial \phi^n$ is the derivative of some scalar function $f(\phi)$.

\subsection{\label{Ssec:DefGravitationalWaves}Isaacson's Definition of Gravitational Waves}

In practice, gravitational waves are usually defined within the framework of linear perturbation theory (see e.g. \cite{misner_gravitation_1973,Stewart:1974uz,WaldBook,Mukhanov:1990me,carroll2019spacetime}) that in the context of Horndeski theory assumes an exact solution to the field equations $\bar{g}_{\mu\nu}$ and $\bar{\phi}$ for some given energy-momentum tensor 
written in some chart. Based on such an exact solution, more complex problems can be studied by considering small deviations as approximate solutions of the form
\begin{equation}
    g_{\mu\nu}=\bar{g}_{\mu\nu}+h_{\mu\nu}\,,\quad\phi=\bar{\phi}+\varphi\,,
\end{equation}
where
\begin{equation}
    |\bar{g}_{\mu\nu}|\,,\; |\bar{\phi}| \sim 1 \,,\quad
    |h_{\mu\nu}|\,,\; |\varphi| \ll 1\,.
\end{equation}
These metric and scalar field perturbations are potentially but not necessarily associated to perturbations in the matter distribution
\begin{equation}
    \Psi_\text{m}=\bar{\Psi}_\text{m}+ \psi_\text{m}\,.
\end{equation}

The diffeomorphism invariance of metric theories defined on a manifold $M$ then translates into a gauge freedom of any perturbed field
\begin{equation}
    \Psi=\bar\Psi+\psi\,,
\end{equation}
on the manifold of the form
\begin{align}\label{eq:GaugeTransf}
 \psi &\rightarrow \psi-\mathscr L_\xi \bar\Psi\,.
\end{align}
with $\mathscr L_\xi$ the Lie derivative along a small vector field $\xi^\mu$. This in particular includes the metric and scalar perturbations $h_{\mu\nu}$ and $\varphi$, for which the gauge transformations become
\begin{align}
    \mathscr L_\xi \bar{g}_{\mu\nu}&=\left(\xi^\alpha\partial_\alpha \bar{g}_{\mu\nu}+2\,\bar{g}_{\alpha(\nu}\partial_{\mu)} \xi^\alpha\right)\,,\label{eq:MetricTransf}\\
   \mathscr L_\xi \bar{\phi}&=\xi^\alpha\partial_\alpha \bar{\phi}\,.\label{eq:ScalarTransf}
\end{align}


The perturbative approach further allows a separation of the field equations [Eq.~\eqref{eq:FE}] into a background equation 
\begin{subequations}\label{eq:FEBackground}
\begin{align}
   \mathcal{G}_{\mu\nu}[\bar{g},\bar{\phi}]&=\bar\kappa\, T_{\mu\nu}[\bar{g},\bar{\Psi}_m]\,,\label{eq:MetricFE background}\\
    \mathcal{J}[\bar{g},\bar{\phi}]&=0\,,\label{eq:{eq:ScalarFE background}}
\end{align}
\end{subequations}
and the equations of motion of the perturbations
\begin{subequations}\label{eq:FEPert}
\begin{align}
    \sum_{i=1}\phantom{}_{\mys{(i)}}\mathcal{G}_{\mu\nu}[h,\varphi]&=\kappa\, \sum_{i=1}\phantom{}_{\mys{(i)}}T_{\mu\nu}[h,\psi_\text{m}]\,,\label{eq:MetricFEPert}\\
    \sum_{i=1}\phantom{}_{\mys{(i)}}\mathcal{J}[h,\varphi]&=0\,,\label{eq:ScalarFEPert}
\end{align}
\end{subequations}
where $\phantom{}_{\mys{(i)}}O[A, B,\dots]$ denotes the $i$th order expansion of the operator $O$ in the corresponding perturbation fields $\{ A,B,\dots \}$, and we omit the explicit dependence on the background.
Choosing the exact solution to be either flat Minkowski space far away from any matter source or a homogeneous and isotropic cosmological spacetime, the above equations allow the study of propagating waves defined as the dynamical solutions of perturbations on top of the given exact solution. 

However, strictly speaking a perturbative description of gravitational waves is in general only well-defined if an additional assumption is made which is often not stated explicitly \cite{Isaacson_PhysRev.166.1263,Isaacson_PhysRev.166.1272,misner_gravitation_1973,Flanagan:2005yc,maggiore2008gravitational,Heisenberg:2023prj}: The definition of gravitational waves propagating on top of a background spacetime requires a clear separation of typical scales of variation, for instance in terms of a low and high-frequency scale $f_L$ and $f_H$ respectively, where
\begin{equation}\label{eq:Separation of Frequencies}
    f_L\ll f_H\,.
\end{equation}
Only such a parametric separation of physical scales leads to an unambiguous notion of dynamical waves propagating on a slowly varying or static background with a meaningful association of a localized energy-momentum of the propagating waves. Observe that for the simplest examples of a fixed Minkowski or cosmological background, such an assumption is implicitly satisfied. 

Yet, the explicit consideration of the so-called Isaacson assumption in Eq.~\eqref{eq:Separation of Frequencies} turns out to be important regardless of the simplicity of the background. This is foremost due to the intrinsic non-linearity of gravitation, implying that any localized energy-momentum content of gravitational waves will inevitably introduce corrections to the low-frequency background, regardless of the exact solution about which one is perturbing. More precisely, a purely linear treatment of gravitational waves fundamentally misses out on an additional non-negligible contribution that in particular in the asymptotically flat limit precisely corresponds to a gravitational wave memory effect \cite{Christodoulou:1991cr,Heisenberg:2023prj}. In this work, we will show that a careful analysis of the Isaacson approach to gravitational waves not only captures such a non-linear memory effect of the gravitational null radiation, but is able to describe both null and ordinary memory of any unbound energy-momentum flux in a unified framework.


In the following, we will therefore assume that any field variable can be split into a dependence on the low- and high-frequency scales, that will be denoted by a corresponding super- or subscript $L$ and $H$. In most cases of interest, the variables of the exact solution about which one is perturbing are either static or slowly varying, such that $\bar{g}_{\mu\nu}=\bar{g}_{\mu\nu}^L$ and $\bar{\phi}=\bar{\phi}^L$. On the other hand, no assumption can be made on the perturbations, which therefore admit a split
\begin{subequations}
\begin{align}
    h_{\mu\nu}&=h_{\mu\nu}^L+h_{\mu\nu}^H\,,\\
    \varphi &= \varphi^L+\varphi^H\,,\\
    \psi_\text{m} &= \psi_\text{m}^L+\psi_\text{m}^H\,.
\end{align}
\end{subequations}

The potential presence of such low-frequency perturbations associated to the slowly-varying background has important consequences. This is foremost due to the fact that accounting for the typical scales of variation introduces the additional small parameter $f_L/f_H\ll 1$, implying that Eqs.~\eqref{eq:FEPert} can in principle not simply be solved order by order in the label $(i)$. 
Moreover, the perturbed equations of motion further admit a split into a low- and high-frequency set of relations. In other words, starting from Eqs.~\eqref{eq:FEPert} there are two distinct sets of equations, whose leading order require a departure from pure linear perturbation theory.

The general strategy for obtaining such a decomposition into low- and high-frequency parts is via a (space-time) average that we will denote by $\langle...\rangle$, whose details are not important as long as it satisfies a list of criteria and in particular allows for integrations by parts of derivative operators. Detailed requirements and properties of the averaging operator can be found in \cite{Stein:2010pn}. The low-frequency contribution of any operator $O$ is therefore given by
\begin{equation}
    \left[O\right]^L=\langle O\rangle\,, \label{eq:lowfreq}
\end{equation}
while the corresponding high-frequency part simply reads 
\begin{equation}
    \left[O\right]^H=O-\langle O\rangle\,.
\end{equation}

\subsection{\label{Ssec:Equations in Asymptotically flat Spacetime}The Leading Order Equations in the Asymptotically Flat Limit}

In this work, we will refrain from a general treatment of the associated equations and will directly analyze the simplest situation by considering the asymptotically flat region only. For more details, we refer to the treatment in \cite{Heisenberg:2023prj}. 

In this case, we assume the existence of a source centered coordinate system $\{t,r,\theta,\phi\}$, where we will often collectively denote the angles as $\Omega=\{\theta,\phi\}$, in which the metric and the scalar field asymptote towards a Minkowski background $\bar{g}_{\mu\nu}=\eta_{\mu\nu}$ and $\bar{\phi}=\text{constant}$ with associated $\mathcal{O}(1/r)$ corrections that naturally represent the perturbed quantities. This requires that at the level of the exact solution the matter background vanishes in the asymptotic limit $\bar{\Psi}_\text{m}=0$, with 
\begin{equation}
    T_{\mu\nu}[\eta,\bar\Psi_m =0]=0\,,
\end{equation}
while asymptotic flatness further requires that any energy-momentum tensor is localized enough such that $\phantom{}_{\mys{(1)}}T_{\mu\nu}=0$. 

Thus, the assumption of an asymptotically flat spacetime naturally selects an exact Minkowski solution $\{\eta_{\mu\nu},\bar{\phi}\}$ that preserves local Lorentz invariance, about which one can consider perturbations associated to the $\mathcal{O}(1/r)$ corrections. We then assume that the bulk motion of a typical source of gravitational waves produces an asymptotic radiation of typical characteristic frequency, setting the scale of $f_H$ of the problem. While any localized source of course also entails static $\mathcal{O}(1/r)$ corrections to the asymptotic fields, these contributions remain non-dynamical to a first approximation and are therefore uninteresting for any gravitational wave experiment \cite{misner_gravitation_1973,Flanagan:2005yc}. In such a situation, any potential presence of propagating low-frequency perturbations therefore only arise at the non-linear level.
We therefore assume the presence of so-called primary waves described by the quantities $h^H_{\mu\nu}$ and $\varphi^H$ and solve Eqs.~\eqref{eq:FEBackground} and \eqref{eq:FEPert}.

In this case, the purely low-frequency equations for the exact Minkowski background [Eqs.~\eqref{eq:FEBackground}] in Horndeski theory become
\begin{subequations}\label{eq:FEFlat}
\begin{align}
    \langle\mathcal{G}_{\mu\nu}[\eta,\bar\phi]\rangle&=\frac{1}{2}\eta_{\mu\nu}G_{2}(\bar{\phi},0)=0\,,\label{eq:MetricFEFlat}\\
    \langle\mathcal{J}[\eta,\bar\phi]\rangle&=G_{2,\phi}(\bar{\phi},0) =0\,.\label{eq:{eq:ScalarFEFlat}}
\end{align}
\end{subequations}
Given our assumptions on the background fields, these background equations translate to two conditions on the functional form of the lowest-order free functionals
\begin{equation}
    \bar{G}_{2}=0\,,\quad \bar{G}_{2,\phi}=0\,,
\end{equation}
where we defined the shorthand notation $\bar{G}_i\equiv G_i(\bar{\phi},0)$.

At the next order, the leading order high-frequency equations within [Eqs.~\eqref{eq:FEPert}] provide a propagation equations for the primary waves that still remains at the linear order in perturbations (see also \cite{Heisenberg:2023prj}). In the case of Horndeski theory, these are of the form
\begin{subequations}\label{eq:FEHighFrequencyProp}
\begin{align}
    \left[\phantom{}_{\mys{(1)}}\mathcal{G}_{\mu\nu}\right]^H&=\phantom{}_{\mys{(1)}}\mathcal{G}_{\mu\nu}[h^H,\varphi^H]=0 \label{eqn:highfreqgHorndeski}\\
    = \bar{G}_{4,\phi}&\,(\partial_\mu \partial_\nu \varphi^H -\eta_{\mu\nu} \Box \varphi^H)-\phantom{}_{\mys{(1)}}G_{\mu\nu}[h^H] \, \bar{G}_4  \,, \nonumber\\
    \left[\phantom{}_{\mys{(1)}}\mathcal{J}\right]^H&=\phantom{}_{\mys{(1)}}\mathcal{J}[h^H,\varphi^H]=0 \label{eqn:highfreqphiHorndeski}\\
    = \varphi^H \bar{G}&_{2,\phi\phi} + \Box \varphi^H (\bar{G}_{2,X} - 2 \bar{G}_{3, \phi}) + \phantom{}_{\mys{(1)}}R[h^H] \bar{G}_{4,\phi} \,,\nonumber
\end{align}
\end{subequations}
where $R[g]$ denotes the Ricci scalar, while 
\begin{equation}
    G_{\mu\nu}[g]\equiv R_{\mu\nu}-\frac{1}{2}g_{\mu\nu}R\,,
\end{equation}
is the Einstein tensor with respect to the metric $g_{\mu\nu}$.

On the other hand, it is at the low-frequency level of the equations of motion where the linear order in high-frequency perturbations needs to be left behind. This is because any contribution involving only one instance of the high-frequency fields averages out to zero on the low-frequency scales, such that only a coarse grained contribution at second order survives that acts as a source for the low frequency background (see also \cite{misner_gravitation_1973,Heisenberg:2023prj}).
The associated leading order low-frequency perturbation equations therefore read
\begin{subequations}\label{eq:FELowFrequency}
\begin{align}
    \phantom{}_{\mys{(1)}}\mathcal{G}_{\mu\nu}[h^L,\varphi^L]&=-\langle\phantom{}_{\mys{(2)}}\mathcal{G}_{\mu\nu}[h^H,\varphi^H]\rangle+\kappa \langle\phantom{}_{\mys{(2)}}T_{\mu\nu}\rangle\,,\label{eq:MetricFELowFrequency}\\
    \phantom{}_{\mys{(1)}}\mathcal{J}[h^L,\varphi^L]&=-\langle\phantom{}_{\mys{(2)}}\mathcal{J}[h^H,\varphi^H]\rangle\,,
\end{align}
\end{subequations}
In the reminder of this section we will however only be interested in considering the high-frequency equations [Eqs.~\eqref{eq:FEHighFrequencyProp}] that describes the propagation of the gravitational waves. The low-frequency equations above giving rise to the memory corrections will then be considered in Sec.~\ref{sec:MemoryHorndeski} below.



\subsection{\label{Ssec:GaugeInv}SVT Decomposition and the Propagation Equation of the Primary Waves}

Let's therefore analyze the leading order propagation equations [Eqs.~\eqref{eq:FEHighFrequencyProp}] more closely. First of all, note that both these linear equations involve both types of high-frequency perturbation fields, which we would like to decouple in order to describe the propagation of individual dynamical degrees of freedom. Moreover, as discussed above [see Eq.~\eqref{eq:GaugeTransf}], the field perturbations only describe physical modes up to gauge transformations. We should therefore both disentangle the field perturbations and identify the propagating physical modes.

A universal way to accomplish both of these tasks is to split the linear perturbation fields in a Helmholtz decomposition within a given asymptotic chart. This is done by first splitting all field perturbations into irreducible parts under the rotation group $SO(3)$, and further decompose the individual pieces into scalar, vector and tensors of the $SO(2)$ subgroup around a given direction. Given the spacial rotational invariance of the background, the resulting scalar, vector and tensor modes in the linear equations of motion will automatically be separated from each other. Moreover, this setup allows the identification of manifestly gauge-invariant, and therefore physical, quantities.

Very generally, the SVT decomposition of $h^H_{\mu\nu}$ as a two index tensor field reads
\begin{subequations}\label{eq:SVTmetricDecomp}
\begin{align}
    h^H_{00}&=2A\,,\\
    h^H_{0i}&=B^\text{T}_i + \partial_i B\,,\\
    h^H_{ij}&=\frac{1}{3} \delta_{ij} D + E^\text{TT}_{ij} + \partial_{(i}E^\text{T}_{j)} + (\partial_i\partial_j -\frac{1}{3} \delta_{ij} \partial^2) E \,,
\end{align}
\end{subequations}
where $D = \delta^{ij}h^H_{ij}$ captures the trace of $h^H_{ij}$, $B^\text{T}_i$ and $E^\text{T}_i$ are transverse vectors such that $\partial^i B^\text{T}_i = 0$ and $\partial^i E^\text{T}_i = 0$
, and $E_{ij}^\text{TT}$ is the transverse-traceless (TT) part of $h^H_{\mu\nu}$, and therefore satisfies $\partial^i  E_{ij}^\text{TT} = \delta^{ij} E_{ij}^\text{TT}=0$.

In such a decomposition it is useful to introduce an explicit spacial orthonormal basis set by a particular radial direction $\hat{n}$. At each given location one can choose two transverse vectors $\hat{u}$ and $\hat{v}$ to $\hat{n}$, such that the basis $\left \{  \hat{n}, \hat{u}, \hat{v} \right \}$ satisfies the completion relation (see e.g. \cite{maggiore2008gravitational,poisson2014gravity})
\begin{equation}
    \delta_{ij} = n_i n_j + u_i u_j + v_i v_j\,.
\end{equation}

With this basis at hand, it is possible to construct the transverse projection operator
\begin{equation}
    P_{i j}(\hat{n}) \equiv \delta_{i j}-n_i n_j=u_i u_j+v_i v_j\,,
\end{equation}
to extract the different components within the SVT decomposition introduced above. For instance, we have
\begin{equation}
   B_i^\text{T}= P_{i j} h_H^{0j}\,.
\end{equation}
Similarly, we define the symmetric transverse-traceless projector
\begin{align}\label{eq:Projector}
    \Lambda_{i j, k l} (\hat{n}) \equiv P_{i k} P_{j l}-\frac{1}{2} P_{i j} P_{k l}\,,
\end{align}
that allows for the extraction of any TT component, for example 
\begin{align}
    E^\text{TT}_{ij} = \Lambda_{i j, k l}h_H^{kl}\,.
\end{align}

While the above decomposition naturally disentangles the scalar, vector and tensor sectors of the equations of motion, of course not all of the SVT components are physical due to the gauge transformations in Eq.~\eqref{eq:GaugeTransf} generated by an infinitesimal high-frequency vector field $\xi^\mu_H$.
Indeed, while for the given asymptotically flat background the scalar perturbation remains invariant under the transformation in Eq.~\eqref{eq:ScalarTransf}
\begin{equation}
    \varphi^H\rightarrow\varphi^H\,,
\end{equation}
the metric perturbation transforms as [Eq.~\eqref{eq:MetricTransf}]
\begin{equation}
    h^H_{\mu\nu} \rightarrow h^H_{\mu\nu} - 2 \partial_{(\mu} \xi^H_{\nu)}\,.
\end{equation}
Thus, only six of the ten modes within the metric decomposition in Eq.~\eqref{eq:SVTmetricDecomp} are physical. 

To explicitly identify these physical modes we also Helmholtz split the gauge transformation into \begin{equation}
    \xi^H_\mu = (\xi_0, \xi^\text{T}_i + \partial_i \xi)\,,
\end{equation}
where $\xi_i$ is transverse, such that the SVT components of $h^H_{\mu\nu}$ transform as
\begin{subequations} \label{eqn:GaugeInvTransf}
\begin{align}
    A & \rightarrow A-\dot{\xi_0} \,, \\
    B^\text{T}_i & \rightarrow B^\text{T}_i-\dot{\xi}^\text{T}_i \,, \\
    B & \rightarrow B-\xi_0-\dot{\xi} \,, \\
    D & \rightarrow D-2 \nabla^2 \xi \,, \\
    E & \rightarrow E -2 \xi \,, \\
    E^\text{T}_i & \rightarrow E^\text{T}_i-2 \xi^\text{T}_i \,, \\
    E_{i j}^\text{TT} & \rightarrow E_{i j}^\text{TT} \,.
\end{align}
\end{subequations}
While therefore $E_{i j}^\text{TT}$ is already gauge invariant, it is not difficult to verify that the following two scalar and two vector combinations represent the additional gauge invariant quantities we seek
\begin{subequations}\label{eqn:GaugeInvQuantity}
\begin{align}
    \Phi & \equiv A-\dot{B}+\frac{1}{2} \ddot{E} \,,\\
    \Theta & \equiv \frac{1}{3}\left(D-\nabla^2 E\right) \,,\\
    \Xi_i & \equiv B^\text{T}_i-\frac{1}{2} \dot{E}^\text{T}_i \,.
\end{align}
\end{subequations}

We have therefore identified the seven physical degrees of freedom of the problem, given by the three gauge invariant scalars $\Phi$, $\Theta$ and $\varphi^H\equiv \varphi$, the two vector modes within $\Xi_i $ together with the two TT tensor modes $E_{i j}^\text{TT}$. It remains to plug this gauge invariant decomposition into the equations of motion of the theory in order to extract the subset of physical modes that are propagating.
As we explicitly show in Appendix~\ref{Appendix:GaugeInvEq}, the linear high-frequency propagation equations within Horndeski theory dictate that, as expected, only the scalar field perturbation $\varphi$ and the TT part of the metric $E_{i j}^\text{TT}$ satisfy dynamical equations of motion
\begin{align}
    \Box E^\text{TT}_{ij} = 0 \,, \quad (\Box-m^2)\varphi = 0 \,. \label{eqn:HorndeskiWaveEq}
\end{align}
with the effective mass of the scalar field perturbation defined as
\begin{align}
    m^2 \equiv \frac{\bar{G}_{2,\phi\phi}}{\bar{G}_{2,X}-2\bar{G}_{3,\phi}+3(\bar{G}_{4,\phi})^2 / \bar{G_4} } \,, \label{eq:defeffmass}
\end{align}
while all other gauge invariant variables are constraints to take the following values
\begin{align}\label{eqn:HorndeskiConstraintEq}
    \Theta &= - \sigma \varphi \,, &
    \Phi &= \frac{1}{2} \sigma \varphi\,, &
    \Xi^i & = 0 \,,
\end{align}
where
\begin{equation}\label{eq:defsigma}
    \sigma \equiv \bar{G}_{4,\phi}/\bar{G}_4 \,.
\end{equation}
 Hence, the linear high-frequency EOMs [Eq.~\eqref{eqn:HorndeskiWaveEq}] indicate that the Horndeski theory has $2+1=3$ \textit{propagating degrees of freedom} reflected in the three independent modes of the high-frequency primary waves, while the remaining four physical modes are constrained, non-dynamical variables.

 The solutions to the homogeneous wave equations in Eqs.~\eqref{eqn:HorndeskiWaveEq} in the large $r$ limit are given in Fourier space by a superposition of plane waves and because of the fixed radial propagation $\hat{n}$ one can indeed without loss of generality consider each plane wave mode individually\footnote{Note that interactions between individual modes would become important when considering stochastic solutions without fixed propagation directions \cite{maggiore2008gravitational}.}, hence
 \begin{subequations}\label{eqn:waveform first}
 \begin{align}
    E^\text{TT}_{ij} (x) &= \frac{1}{r} C^\text{TT}_{ij}(\Omega) e^{ik_\mu x^\mu} \,,\\ \varphi(x) &= \frac{1}{r} C(\Omega) e^{ip_\mu x^\mu} \,, 
\end{align}
 \end{subequations}
with
\begin{equation}
   k^i\, C^\text{TT}_{ij}=0\,,\quad \delta^{ij} C^\text{TT}_{ij}=0\,,
\end{equation}
and where the perturbations are characterized by their $\mathcal{O}(1/r)$ fall-off. Note that in terms of the Fourier momenta, the direction of propagation is given by
\begin{equation}
      n_i=k_i/k=p_i/p=\partial_i r\,.
\end{equation}
Moreover, Eqs.~\eqref{eqn:HorndeskiWaveEq} imply the following dispersion relations
\begin{subequations}
\begin{align}
    k^0(k)&=c\,k \,, \\
    p^0(p)&=c\,\sqrt{p^2+m^2} \,, \label{eqn:wavevectorHordenski}
\end{align}
\end{subequations}
where we have defined $k\equiv \sqrt{k_ik^i}$ and $p\equiv \sqrt{p_ip^i}$. 
This directly implies that the TT perturbations propagate with the speed of light $c$, while the scalar field perturbation propagates with a group velocity of 
\begin{align}
    v \equiv \frac{\partial p^0(p)}{\partial p}  = \,\sqrt{1-\frac{m^2c^2}{(p^0)^2}} \,. \label{eqn:phivelocity}
\end{align}
This spacial velocity is to be understood as the velocity in the chosen asymptotic source centered frame and is assumed to be non-zero. 

Therefore, the solutions of the three propagating degrees of freedom in Eq.~\eqref{eqn:waveform first} can be written as
\begin{subequations}\label{eqn:waveform second}
\begin{align}
    E^\text{TT}_{ij} (x) &= \frac{1}{r} C^\text{TT}_{ij}(\Omega) \,e^{-ik^0(ct-r)} \,,\\ \varphi(x) &= \frac{1}{r} C(\Omega) \,e^{-ip^0(ct-\beta r)} \,,
\end{align}
\end{subequations}
where we have defined $\beta\equiv v/c$. In other words, the asymptotic solutions of the high-frequency wave modes take on the following general form
\begin{subequations}\label{eqn:waveform}
\begin{align}
    E^\text{TT}_{ij} (x) &= \frac{1}{r} f^\text{TT}_{ij}(ct-r,\Omega) \,,\\ \varphi(x) &= \frac{1}{r} f(ct-\beta r,\Omega) \,.
\end{align}
\end{subequations}
for some transverse-traceless and scalar functions $f^\text{TT}_{ij}$ and $f$.
Observe that these universal asymptotic forms imply that to the leading order in $1/r$, the spatial derivatives and the temporal derivatives of $E^\text{TT}_{ij}$ are related by
\begin{equation}\label{eq:identity derivatives TT}
    \partial_{i} E_\text{TT}^{ij} = -\frac{1}{c} n_i\, \partial_0E_\text{TT}^{ij}\,,
\end{equation}
while similarly for the massive scalar mode we have 
\begin{equation}\label{eq:identity derivatives}
    \partial_{i} \varphi = -\frac{\beta}{c}\, n_i\, \partial_0\varphi\,.
\end{equation}
For simplicity, we set $c=1$ from now on.

\section{\label{Sec:Polarizations}Gravitational Polarization Modes}

As an interlude, we want to offer in this section a self-consistent account of the gravitational polarization modes present in Horndeski theory. While propagating degrees of freedom are defined as the linearly independent gauge invariant solutions to the equations of motion of the leading-order perturbed fields, we define here \textit{gravitational polarization modes} to represent the detectable modes of the physical metric that couples minimally to the matter. When discussing memory beyond GR in the subsequent Section~\ref{sec:MemoryHorndeski} a clear distinction between the concepts of propagating degrees of freedom of the previous section and gravitational polarizations will be essential. Indeed, the geodesic deviation equation at the basis of the definition of gravitational polarization will set the stage for the very definition of memory, as well as a classification of different types of displacement memory that might be present in beyond GR theories. Furthermore, while the polarization content of Horndeski theory was already discussed for instance in \cite{Hou:2017bqj} it is useful to rederive these results within the manifestly gauge-invariant SVT decomposition approach introduced in Sec.~\ref{Ssec:GaugeInv}. To start, we should therefore discuss the concept of additional polarizations in general metric theories beyond GR.



\subsection{Additional Polarizations Beyond GR}
The foundation of any light-travel time based measurement of gravitational waves is given by the geodesic deviation equation, which for two time-like geodesics with spacial proper distance separation vector $s_i$ is to leading-order governed by the electric part of the Riemann tensor (see e.g. \cite{misner_gravitation_1973,maggiore2008gravitational,carroll2019spacetime})
\begin{align}
    \ddot{s}_i=-R_{i 0 j 0} s_j \, , \label{eq:geodesicdeviation}
\end{align}
where $R_{\mu\nu\rho\sigma}$ represents the Riemann tensor of the physical metric $g_{\mu\nu}$.
Note that this equation solely follows from the minimal and universal coupling requirement and is therefore valid in any metric theory of gravity, regardless of the precise form of the equations of motion. Since the electric part of the Riemann tensor possesses six independent components, we can already conclude at this stage, that the physical response to gravitational waves is in principle associated to six distinct gravitational polarization modes. Naturally, these polarization modes are associated to the six gauge invariant degrees of freedom within the perturbations of the physical metric that we identified in Sec.~\ref{Ssec:GaugeInv}.

To explicitly derive this association within the asymptotically flat setup employed in this work, we are interested in computing the leading order contribution to the Riemann tensor of the physical metric
\begin{align}
    R_{\mu\nu\rho\sigma}=\phantom{}_{\myst(1)}R_{\mu\nu\rho\sigma}+\mathcal{O}\left(\frac{1}{r^2}\right)\,,
\end{align}
where the leading order term on the Minkowski background associated to a metric perturbation $h_{\mu\nu}$ reads
\begin{align}
    \phantom{}_{\myst{(1)}}R_{i 0 j 0}=-\frac{1}{2}\left(\partial_0 \partial_0 h_{i j}+\partial_i \partial_j h_{00}-\partial_0 \partial_i h_{0 j}-\partial_0 \partial_j h_{0 i}\right) . \label{eqn:leadingorderRelectric}
\end{align}
At this point it should be noted that the metric perturbation $h_{\mu\nu}$ in principle represents here both the low- and high-frequency dynamical perturbations. However, the SVT decomposition together with the identification of gauge invariant modes explicitly performed in Sec.~\ref{Ssec:GaugeInv} for the high-frequency modes can also be applied in exactly the same manner to the low-frequency components. Thus, in this section, all perturbative quantities are to be understood as the full perturbations that also include potential dynamical low-frequency components. We will however refrain from introducing new symbols for total perturbations in order to keep the notation simple.


With this comment out of the way, we can rewrite the general expression in Eq.~\eqref{eqn:leadingorderRelectric} in terms of the gauge invariant quantities within a general SVT decomposition and obtain
\begin{align}\label{eq:Gauge invariant Riemann first}
    \phantom{}_{\myst{(1)}}R_{0i0j} = \frac{-1}{2} \partial_0 \partial_0 E_{ij}^\text{TT} - \frac{\delta_{ij} }{2} \partial_0 \partial_0 \Theta + \partial_0 \partial_{(i} \Xi_{j)} -\partial_i \partial_j \Phi .
\end{align}
As expected, the gauge invariant linear perturbation of the Riemann tensor can be written entirely in terms of gauge invariant variables of the metric. While therefore all six gauge-invariant modes of the physical metric can in general be involved in the physical response to a gravitational wave, not all of them must be excited. Indeed, for GR for instance, only the two TT tensor modes are present. 

In general, only the components associated to a dynamical mode contribute to the detector response
. However, in order to contribute, the different components of the physical metric must not be dynamical themselves, but it suffices if the gauge invariant variables are associated to a dynamical mode in the gravitational sector. This is for instance the case for the two scalar components $\Theta$ and $\Phi$ that are excited through their non-trivial relation with the dynamical degree of freedom described by $\varphi$ in Eq.~\eqref{eqn:HorndeskiConstraintEq}. To continue the analysis we have to make use of this fact and assume that any mode that contributes must be associated to a dynamical behavior and therefore assume an asymptotic form similar to Eq.~\eqref{eqn:waveform} together with the asymptotic relation Eq.~\eqref{eq:identity derivatives} for some velocity $\beta$ of the associated plane-wave mode (see also \cite{poisson2014gravity}).

This allows us to rewrite Eq.~\eqref{eq:Gauge invariant Riemann first} as
\begin{align}
    \phantom{}_{\myst{(1)}}R_{0i0j} = -\frac{1}{2} \partial_0 \partial_0 A_{ij} \label{eqn:Apolarization}
\end{align}
where the response matrix is given by
\begin{equation}\label{eq:Polarization Matrix}
    A_{ij}\equiv  E_{ij}^\text{TT} + \delta_{ij} \Theta + 2\beta n_{(i} \Xi_{j)} + 2\beta^2 n_i n_j \Phi\,.
\end{equation}
With this equation at hand, the geodesic deviation equation [Eq.~\eqref{eq:geodesicdeviation}] is easily integrated to give to leading order
\begin{align}\label{eq:integratedGeodesicDeviation}
    \Delta s_i (\tau) = \frac{1}{2} \Delta A_{ij}(\tau) s^j (\tau_0) \,,
\end{align}
where
\begin{equation}
    \Delta s_i (\tau)\equiv s_i (\tau)-s_i (\tau_0)\,,
\end{equation}
for some initial reference proper time $\tau_0$ before the presence of any gravitational wave.

The response matrix is therefore what ultimately governs the physical stretch in proper distances within an idealized detector, and is thus also known as the polarization matrix. The six independent components of the polarization matrix can conveniently be expanded in a specific polarization basis as
\begin{align}
    A_{ij} \equiv & P_+ \,e^+_{ij} + P_\times\, e^\times_{ij} + P_u\, e^u_{ij} + P_v \,e^v_{ij}+ P_b\, e^b_{ij} + P_l\, e^l_{ij}  \,,\label{eq:PolM 2}
\end{align}
where
\begin{subequations}\label{eq:PolTensors}
\begin{align}
e^+_{ij}&\equiv u_iu_j-v_iv_j\,,\quad e^\times_{ij}\equiv u_iv_j+v_iu_j\,, \label{eqn:crossandplusdef}\\
e^u_{ij}&\equiv n_iu_j+u_in_j\,,\quad e^v_{ij}\equiv n_iv_j+v_in_j\,,\\
\quad e^b_{ij}&\equiv u_iu_j+v_iv_j\,,\quad e^l_{ij} \equiv n_in_j\,,
\end{align}
\end{subequations}
For concreteness, it is useful to align the direction of travel $\hat{n}$ with $z$-axis, in which case the polarization matrix becomes
\begin{align}
    A_{ij} = \begin{bmatrix}
  P_+ + P_b   & P_\times & P_u \\
  P_\times & -P_+ +P_b & P_v \\
  P_u & P_v & P_l
    \end{bmatrix},
\end{align}
The spin-weighted scalars $P_A$ are ultimately what we define as the gravitational polarizations. Of course $P_+$ and $P_\times$ are the two familiar tensor polarization modes that show up in standard GR, while $P_u$ and $P_v$ represent two additional vector modes, $P_l$ is the scalar longitudinal mode, and $P_b$ is known as the breathing polarization.

Comparing the above expressions in Eq.~\eqref{eq:PolM 2} with Eq.~\eqref{eq:Polarization Matrix} while utilizing the knowledge that $n_i$, $u_i$, $v_i$ form an orthonormal basis, we finally obtain the following relations between the gravitational polarizations and the gauge invariant components of the physical metric perturbations
\begin{subequations}\label{eq:Polarization Modes}
\begin{align}
P_+  &=  \frac{1}{2}E^\text{TT}_{ij}e^{ij}_+\,, & P_u  &= \beta u^i \Xi_i\,, & P_b &= \Theta\,, \\
P_\times  &= \frac{1}{2}E^\text{TT}_{ij}e^{ij}_\times\,, & P_v  &= \beta v^i \Xi_i\,, & P_l  &= \Theta + 2 \beta^2 \Phi\,.
\end{align}
\end{subequations}

\subsection{Gravitational Polarizations in Horndeski Theory}\label{Sec:GravPolHorndeski}
Based on the general considerations above, it is now relatively straightforward to assess the polarization content within Horndeski theory. First, since $\Xi^i = 0$ [Eq.~\eqref{eqn:HorndeskiConstraintEq}] Horndeski theory has no additional vector polarization modes, hence $P_u = P_v = 0$. However, on top of the two tensor polarizations $P_+$ and $P_\times$, Horndeski theory offers the possibility for the presence of additional scalar gravitational polarizations that might be excited through the non-minimal coupling between the additional scalar degree of freedom and the physical metric. Concretely, through Eqs.~\eqref{eqn:HorndeskiConstraintEq} and \eqref{eq:Polarization Modes} we have that
\begin{align}
    P_b= -\sigma \varphi\,,\qquad P_l =\sigma \varphi (-1 +\beta^2)\,.
\end{align}
Therefore, depending on the parameters of the theory, up to four gravitational polarizations may be present within Horndeski gravity. As can easily be seen, the breathing polarization is always present for a nontrivial non-minimal coupling parameter $\sigma$ defined back in Eq.~\eqref{eq:defsigma}, whereas the longitudinal gravitational polarization has the additional requirement of $\beta\neq 1$, and is therefore only present for massive modes, where the mass $m$ is given by Eq.~\eqref{eq:defeffmass}. Note, however, that the two scalar polarizations are not independent of each other, as they are excited by the same dynamical gravitational degree of freedom in the theory.

It is particularly illuminating to illustrate the different possibilities within Horndeski theory by considering the concrete known theories beyond GR that are captured within its framework, introduced at the end of Sec.~\ref{Ssec:Horndeski}. First of all, for Brans-Dicke theory [Eq.~\eqref{eq:BD}] we have that $\sigma=1/\bar\phi\neq 0$, and therefore BD gravity does excite an additional scalar gravitational polarization. However, since the scalar perturbations remains massless $m=0$, the speed of propagation of the wave is always luminal, hence \textcolor{blue}{$\beta=1$}, such that only the breathing mode is present. On the other hand, for $f(R)$ gravity, the background equations require that $\bar\phi=f(\bar\phi)=0$, while however $f'(\phi)$ and $f''(\phi)$ are non-zero. This implies that the theory comes with parameters $\sigma=f''(\bar\phi)/f'(\bar\phi)\neq 0$ and $m^2=f'(\bar\phi)/(3f''(\bar\phi))\neq 0$ and thus excites both $P_b$ and $P_l$. Finally, sGB gravity has $\sigma=0$ and therefore does not come with any additional gravitational polarizations. These examples nicely illustrate the difference between the concepts of propagating degrees of freedom and gravitational polarizations. While all three theories above possess three propagating DOFs in the gravity sector, the scalar mode can either excite none, one or even two additional gravitational polarizations.

\section{\label{sec:MemoryHorndeski}Ordinary and Null Memory in Horndeski Theory}

The gravitational wave memory effect is defined as a permanent displacement between test masses in an idealized detector after the passage of a burst of gravitational waves. More formally, within the solution to the geodesic deviation equation in Eq.~\eqref{eq:integratedGeodesicDeviation}, gravitational memory is defined as a non-zero value of the leading order spacial proper distance displacement between two timelike geodesics in the limit of the absence of any gravitational wave disturbance
\begin{equation}
    \lim_{\tau\rightarrow\infty}\Delta s(\tau)\neq 0\,.
\end{equation}
Therefore, the memory effect relies on a non-zero difference before and after a gravitational wave within one of the components of the polarization matrix 
\begin{equation}
    \lim_{\tau\rightarrow\infty}\Delta A_{ij}(\tau)\neq 0
\end{equation}
and can therefore through Eq.~\eqref{eq:PolM 2} naturally be associate to one of the gravitational polarization types. 

Since, as discussed in Sec.~\ref{Sec:GravPolHorndeski}, in Horndeski theory up to four gravitational polarizations can be excited, gravitational memory can in principle arise both within the tensor and the scalar sector. Concretely, through Eqs.~\ref{eq:Polarization Modes} memory in Horndeski theory can in principle be associated to a non-zero permanent offset within either one of the following gauge invariant variables
\begin{align}
    \Delta A_{ij} = &\frac{1}{2}\Delta E^\text{TT}_{kl}\left(e_+^{lk}\, e^+_{ij} +e_\times^{lk}\, e^\times_{ij} \right)   \nonumber \\
    & + \Delta \Theta ( e^b_{ij} + e^l_{ij}) + \Delta \Phi (2 \beta^2 e^l_{ij}) \,.
\end{align}
In this work, we will however entirely focus on the observationally relevant tensor memory within the TT tensor polarizations.

At first sight, the presence of such a non-zero permanent offset is not guaranteed. Indeed, in Sec.~\ref{Ssec:GaugeInv} we have elaborated on the asymptotic solution of the propagating high-frequency perturbations given by a superposition of plane waves. Very generally, these solutions do not give rise to any memory contribution, as the high frequency part of the gauge invariant variables are subject to symmetric attenuation effects that die out any signal with zero displacement
\begin{equation}
    \lim_{\tau\rightarrow\infty}[\Delta E^\text{TT}_{ij}]^H=0\,.
\end{equation}
However, as we will now show explicitly, the solution of the asymptotic leading order the low-frequency equations of motion [Eq.~\eqref{eq:MetricFELowFrequency}] will give rise to a propagating low-frequency correction to the background spacetime, which precisely provides a memory effect in the sense defined above.

In the following, we will therefore first derive the precise form of the low-frequency Isaacson equation in Horndeski theory within the SVT decomposition introduced in Sec.~\ref{Ssec:GaugeInv}, and solve for the resulting memory evolution that is sourced by the back-reaction of a massive or massless field perturbations in the limit to null infinity.
For simplicity, we will from now on denote the low-frequency part of the metric perturbation as
\begin{equation}
     [h^L_{ij}]^\text{TT}=[E^L_{ij}]^\text{TT}=\delta h^\text{TT}_{ij}\,,
\end{equation}
while the high-frequency waves are simply given
\begin{equation}
     [h^H_{ij}]^\text{TT}=[E^H_{ij}]^\text{TT}=h_{ij}^\text{TT}\,,\quad \varphi^H=\varphi\,.
\end{equation}


\subsection{\label{Ssec:SolveMemoryEq} Memory Evolution Equation in Horndeski Theory}

We have obtained and solved the linear propagating equations in the previous parts. Let us now consider the low-frequency back-reaction equation Eq.~\eqref{eq:FELowFrequency} in the asymptotic limit, which provides us with the memory evolution equation in Horndeski gravity. Focusing on the tensor memory, we will however only consider the metric equations of motion and single out the TT part of the equation by the use of the projection operator introduced in Eq.~\eqref{eq:Projector}.

First, observe that the left-hand side of this equation takes the same form as the high-frequency metric field equation. Hence, by performing the same 3+1 and SVT decomposition and grouping the perturbation fields into gauge-invariant quantities as in Sec.~\ref{Ssec:GaugeInv}, the dynamical part of the left-hand side of the back-reaction equation becomes
\begin{align}
     \left ( \phantom{}_{\mys{(1)}}\mathcal{G}_{ij}[h^L,\varphi^L] \right )^\text{TT}= \frac{-1}{2} \Bar{G}_4 \Box \delta h^\text{TT}_{ij} \,.
\end{align} 
The dynamical TT part of the low frequency equation therefore becomes
\begin{align}\label{eq:intermediate low f}
    \frac{-1}{2} \Bar{G}_4 \Box \delta h^\text{TT}_{ij} = \Lambda_{ij,kl} (-\langle\phantom{}_{\mys{(2)}}\mathcal{G}^{kl}[h^H,\varphi^H]\rangle+\kappa \langle\phantom{}_{\mys{(2)}}T^{kl}\rangle) \,.
\end{align}
Using the SVT decomposition, together with the properties of the spacetime averaging [Sec.~\ref{Ssec:DefGravitationalWaves}] that allows for integrations by parts, as well as the leading propagation equation of the high-frequency fields, the first term of the right-hand side can then be evaluated to give
\begin{align}\label{eq:intermediate RHS low f}
    \langle\phantom{}_{\mys{(2)}}\mathcal{G}_{ij}\rangle = -\frac{1}{4}\bar{G}_4 \left \langle \partial_i h^\text{TT}_{kl} \partial_j h_\text{TT}^{kl} \right \rangle - \frac{\bar{G}_{2,\phi\phi}}{2 m^2} \left \langle \partial_i \varphi \partial_j \varphi \right \rangle \,.
\end{align}

Thus, the TT part of the leading order low-frequency equation [Eq.~\eqref{eq:intermediate low f}] in Horndeski theory can be written as
\begin{align}\label{eq:memory equation}
    \Box \delta h^\text{TT}_{ij} = - 2\Lambda_{ij,kl} \left(\kappa^\text{($h$)} t^{kl}_\text{($h$)} + \kappa^\text{($\varphi$)}t^{kl}_\text{($\varphi$)} 
     +\kappa^\text{(m)} t^{kl}_\text{(m)} \right)\,,
\end{align}
where
\begin{subequations}\label{eq:HornEMTs}
\begin{align}
    t^{kl}_\text{($h$)} &= \left \langle \partial^k h^{TT}_{kl} \partial^l h_{TT}^{kl} \right \rangle\,,\label{eq:h HornEMTs}\\
    t^{kl}_\text{($\varphi$)} &= \left \langle \partial^k \varphi \partial^l \varphi \right \rangle\,,\label{eq:phi HornEMTs}\\
    t^{kl}_\text{(m)}&=  \langle\phantom{}_{\mys{(2)}}T^{kl}\rangle\,,\label{eq:m HornEMTs}
\end{align}
\end{subequations}
represent the effective energy-momentum tensors of the high-frequency waves as well as the unbound matter components, and the coefficients read
\begin{subequations}\label{eq:coeff}
\begin{align}
    \kappa^\text{($h$)} &= \frac{1}{4} \,,\label{eq:kappa h} \\
    \kappa^\text{($\varphi$)} &= \frac{\bar{G}_{2,X}-2\bar{G}_{3,\phi}}{2\bar G_4}+\frac{3}{2}\sigma^2 = \frac{\bar{G}_{2,\phi\phi}}{2\bar G_4m^2} \,, \label{eq:kappa phi}\\
    \kappa^\text{(m)} &= \frac{\kappa}{\bar{G}_4}\label{eq:kappa m}\,.
\end{align}
\end{subequations}
The solution of the memory evolution will then be a superposition of the solutions of inhomogeneous wave equations of the three different types of asymptotic energy-momentum sources.

\subsection{\label{Ssec:FormOfAsymptoticEM} Asymptotic Energy-Momentum Fluxes}

Next we want to show that any asymptotic, radial energy-momentum tensor is entirely characterized by its 00-component, hence the energy density. More precisely, any energy-momentum flux with constant radial velocity $v$ in the radiation zone takes on the general form in Eq.~\eqref{eq:General Form asymptotic EMT}.

Before we give the general argument, we want to explicitly verify this statement in the present specific case of Horndeski theory through direct computation.
For this, we first present also the 00 component of the second-order metric equations of motion of Horndeski gravity within the SVT decomposition
\begin{equation}
    \langle\phantom{}_{\mys{(2)}}\mathcal{G}_{00}\rangle = -\frac{1}{4}\bar{G}_4 \left \langle \partial_0 h^\text{TT}_{kl} \partial_0 h_\text{TT}^{kl}\right \rangle - \frac{\bar{G}_{2,\phi\phi}}{2 m^2} \left \langle \partial_0 \varphi \partial_0 \varphi \right \rangle \,,
\end{equation}
and compare the result to Eq.~\eqref{eq:intermediate RHS low f}, such that
\begin{subequations}
\begin{align}
    t_{00}^\text{($h$)} &= \left \langle \partial_0 h^{TT}_{kl} \partial_0 h_{TT}^{kl} \right \rangle\,,\label{eq:T00 h}\\
    t_{00}^\text{($\varphi$)} &= \left \langle \partial_0 \varphi \partial_0 \varphi \right \rangle\,.\label{eq:T00 phi}
\end{align}
\end{subequations}

First of all, as in GR, the massless high-frequency wave satisfies the asymptotic relation given in Eq.~\eqref{eq:identity derivatives TT}, which implies that
\begin{align}
    t_{ij}^{(h)} =  n_i n_j \left \langle \partial_{0} h^\text{TT}_{kl} \partial_{0} h_\text{TT}^{kl} \right\rangle =  n_i n_j t_{00}^{(h)} \,.\label{eq:EMT h 00}
\end{align}
This precisely corresponds to Eq.~\eqref{eq:General Form asymptotic EMT} with the luminal speed of $v=1$.
Similarly, the potentially massive field wave $\varphi$ satisfies Eq.~\eqref{eq:identity derivatives}, such that
\begin{align}
    t_{ij}^\text{($\varphi$)} = \beta^2 n_i n_j \left \langle \partial_{0} \varphi \partial_{0} \varphi \right \rangle = \beta^2 n_i n_j t_{00}^\text{($\varphi$)} \,. \label{eq:EMT phi 00}
\end{align}

As an example for the asymptotic energy-momentum tensor of unbound matter, that we will later use, consider also the case of a freely moving particle of mass $M$ that is gravitationally unbound from other particles. In this case, the asymptotic energy-momentum tensor of matter takes the form (see e.g. \cite{Weinberg1972})
\begin{align}
    T_{\mu\nu}^\text{(m)}(x) = M\gamma V_\mu V_\nu \delta^{(3)}(\vec{x}-\vec{x}'(t))  \, , \label{eq:mattersourcesimp}
\end{align}
with velocity vector $V_\mu = (-1, \vec{\beta})$ and where $x'(t)$ denotes the radial trajectory of the particle.
Hence, component-wise we can write $V_i = -\beta n_i V_0$, and then
\begin{align}
   T_{ij}^\text{(m)}(x) = \beta^2 n_i n_j T_{00}^\text{(m)}(x) \,.
\end{align}
A multi-particle system can be considered as the sum of all single-particle systems.

Thus, indeed, any of the three asymptotic energy momentum tensors $T^\text{a}_{\mu\nu}$ described above has the special form
\begin{equation}\label{eq:General Form asymptotic EMT}
    T^\text{a}_{ij}(u,r,\Omega)=T^\text{a}_{00}\,\beta^2 \,n_in_j\,, 
\end{equation}
with $\beta=1$ for luminal fluxes.
As mentioned, this is a very general result that holds for any asymptotic energy-momentum tensor $T^\text{a}_{\mu\nu}$ that carries away energy from the source at some constant asymptotic radial velocity $v$.

To derive this general result, note first that by definition the energy density $T^\text{a}_{00}$ of an energy flux with constant asymptotic velocity ratio $\beta$ defines the energy flux
\begin{equation}\label{eq:energy flux}
    \beta T^\text{a}_{00}(t,r,\Omega)=\frac{1}{r^2}\frac{dE(t-r/\beta,\Omega)}{dtd\Omega}\,.
\end{equation}
Hence, the energy flux only depends on the angles $\Omega$ and the 
asymptotic retarded time
\begin{equation}
    u=t-\frac{r}{\beta}\,.
\end{equation}
As a consequence of the dependence on the specific combination of asymptotic retarded time, to leading order in $1/r$, the energy density satisfies
\begin{equation}\label{eq:IdentityEMT deriv prime}
    \partial_i T_\text{a}^{00}=-\frac{n_i}{\beta}\partial_0 T_\text{a}^{00}\,,
\end{equation}
such that
\begin{equation}\label{eq:IdentityEMT deriv}
    \partial_0 T_\text{a}^{00}=-\beta \,n^i\partial_i T_\text{a}^{00}\,.
\end{equation}

On the other hand, the time component of the conservation of the energy-momentum tensor in the radiation zone
\begin{equation}\label{eq:Conservation asymptotic EMT}
    \partial_\mu T_\text{a}^{\mu\nu}=0\,,
\end{equation}
also implies that 
\begin{equation}\label{eq:relation partial i EMT}
    \partial_i T_\text{a}^{i0}=-\partial_0 T_\text{a}^{00}=\beta^i\,\partial_i T_\text{a}^{00}\,,
\end{equation}
where in the last equality we have employed Eq.~\eqref{eq:IdentityEMT deriv}. Assuming that we are only interested in the radial components of the energy momentum tensor, one can integrate this equation over a spherical shell in the radiation zone of volume element $dV=dr r^2 d\Omega$ and reduce the integral through Stocke's theorem to the outer surface with unit normal $n_i$ (see also \cite{maggiore2008gravitational}) to obtain 
\begin{equation}\label{eq:RelTi0 init}
    \int_{S^2}   d\Omega\, n_i\,T_\text{a}^{i0}= \int_{S^2}   d\Omega\, \beta \,T_\text{a}^{00}\,.
\end{equation}
Up to transverse terms that we are neglecting, we thus have the relation
\begin{equation}\label{eq:RelTi0}
    T_\text{a}^{i0}=\beta \,n^i \,T_\text{a}^{00}\,.
\end{equation}
Similarly, the spacial components of Eq.~\eqref{eq:Conservation asymptotic EMT} require
\begin{equation}
    \partial_i T_\text{a}^{ij}=-\partial_0 T_\text{a}^{0j}=-\beta n^j \,\partial_0 T_\text{a}^{00}= \beta^2\,n^in^j\,\partial_i T_\text{a}^{00}\,,
\end{equation}
such that using similar arguments as above, up to transverse terms one indeed recovers the relation
\begin{equation}\label{eq:RelTij}
    T_\text{a}^{ij}=\beta^2\,n^i n^j \,T_\text{a}^{00}\,.
\end{equation}


\subsection{A Unified Treatment of Null and Ordinary Memory}

Given the general considerations of asymptotic energy momentum tensors discussed above, we will now be able to provide a very general form of the solution for the low-frequency memory evolution equation that reads [Eq.~\eqref{eq:memory equation}]
\begin{align}\label{eq:backreaction}
    \Box \delta h^\text{TT}_{ij} = -2\kappa_{\myst{eff}} \Lambda_{ij,kl} T_\text{a}^{kl} \,. 
\end{align}
As discussed, in the case of Horndeski theory, the energy-momentum tensor $T_\text{a}^{kl}$ can either be the one associated to the tensor waves $t_\text{(h)}^{kl}$, the gravitational scalar field perturbation $t_\text{($\varphi$)}^{kl}$ or the unbound matter $T^{kl}_\text{(m)}$. However, the memory evolution equation can be solved in the same way regardless of the type of conserved energy-momentum source, as long as it satisfies the asymptotic form in Eq.~\eqref{eq:General Form asymptotic EMT}. The steps necessary to obtain the general solution are very close to what was already carried out for instance in \cite{PhysRevD.44.R2945,Garfinkle:2022dnm}, although the underlying context within these works is rather different. Moreover, our treatment offers a generalization to wave sources associated to massive field perturbations with asymptotic speeds different from $c$.

First, a general solution to Eq.~\eqref{eq:backreaction} is of course provided by introducing a standard Green's function $G(x-x')$ that satisfies 
\begin{align}
    \Box_{x} G(x-x') = \delta^{(4)} (x-x') \,,
\end{align}
with $\Box_{x}$ being the d'Alembert operator with respect the variable $x$.
The corresponding solution to Eq.~\eqref{eq:backreaction} is then given by
\begin{align}
    \delta h^\text{TT}_{ij} (x) & =-2\kappa_{\myst{eff}} \,\Lambda_{ij,kl}  \int d^{4}x' \; G(x-x') \, T_\text{a}^{kl} (x')   \label{eqn:SolForm} \,.
\end{align}
Here the appropriate Green's function is given by (see e.g. \cite{maggiore2008gravitational})
\begin{align}
    G(x-x') = -\frac{\delta (x^{0}_{\text{ret}}-x'^{0})}{4\pi \left | \vec{x}-\vec{x}' \right | }  \, ,
\end{align}
where $x'^{0}=t'$, $x^{0}_{ret}=t_\text{ret}$, and $t_\text{ret}$ is the retarded time for the propagating gravitational wave,
\begin{align}
    t_\text{ret}= t- \left | \vec{x}-\vec{x}' \right | \, .
\end{align}

However, we want to evaluate the memory within the massless TT polarization mode travelling at the speed of light in the limit to null infinity outside its own source. This limit to null infinity is provided by the large $r$ limit, but at fixed asymptotic retarded time $u=t-r/c$. We should therefore transform the coordinates to
\begin{equation}
    (t,x,y,z) \rightarrow (u,r,\theta,\phi)\,.
\end{equation}
Note, however, that as before we will only change coordinates in the arguments of the fields while treating the indices still in source-centered Minkowski coordinates.

Similarly, given the special form of the asymptotic energy flux in Eq.~\eqref{eq:General Form asymptotic EMT} together with Eq.~\eqref{eq:energy flux} of asymptotic radial speed $v$ equal or less than $c$, it is convenient to perform the transformation
\begin{equation}
    (t',x',y',z') \rightarrow (u',r',\theta',\phi')\,,
\end{equation}
of the source coordinates, where in this case the asymptotic retarded time reads $u'=t'-r'/v$. Moreover, the volume form in these new coordinates becomes
\begin{align}
    d^{4}x' \rightarrow r'^{2}du'dr'd\Omega' \, .
\end{align}

Under these changes of coordinates, we can also of course also write $\Vec{x}=r\Vec{n}$, $\Vec{x}'=r'\Vec{n}'$, where $\Vec{n}$ provides the direction of the observer with respect to the coordinate origin, while $\Vec{n}'$ represents the direction from the same origin, at which the unbounded source creating the back-reaction is evaluated. Note that as asymptotic radial directions, both $\Vec{n}$ and $\Vec{n}'$ correspond to the asymptotic longitudinal direction already employed above. 

It is important to realize, however, that while the source terms of the memory equation in Eq.~\eqref{eq:backreaction} are non-zero in the asymptotic region, there still exists an effective hierarchy between $r$ and $r'$ as the memory component evaluated at $x$ only depends on the unbounded source terms for which $r' \ll r$ holds (see \cite{Garfinkle:2022dnm,Heisenberg:2023prj}).
This allows us to expand the source-to-observer vector in the usual way as
\begin{align}
    \left | \vec{x}-\vec{x}' \right | &= \sqrt{(\vec{x}-\vec{x}')^2}  \nonumber \\
    &=\sqrt{r^2+r'^2-2rr'\Vec{n} \cdot \Vec{n}'} \nonumber \\
    &\simeq r-r'\Vec{n}' \cdot \Vec{n} \, . 
\end{align}

Moreover, under this approximation and by transforming to the asymptotic retarded time coordinates the Green's function can be simplified to the following form
\begin{align}
    G(x-x') &=  -\frac{ \delta ((u-u')+r'(\Vec{n}' \cdot \Vec{n}-1/\beta)) }{4\pi (r-r'\Vec{n}' \cdot \Vec{n})} \nonumber \\
    &=  -\frac{\beta\,\mathcal{V}\,\delta (r'-v(u-u')\mathcal{V})  }{4\pi (r-r'\Vec{n}' \cdot \Vec{n})}\, .
\end{align}
where we have defined
\begin{equation}
    \mathcal{V}\equiv\frac{1}{1-\beta \Vec{n}' \cdot \Vec{n}}
\end{equation}
The last step was obtained by the formula 
\begin{equation}
    \delta(g(x))= {\textstyle \sum_{i}} \delta(x-x_{i})/ \left | g'(x_i) \right |\,,
\end{equation}
where $x_i$'s are roots of $g(x)$.

Gathering all the ingredients above while killing the $r'$ integral with the $\delta$-function, the gravitational wave memory in Eq.~\eqref{eqn:SolForm} becomes
\begin{align}
    \delta h_{ij}^\text{TT} (x) = \frac{\kappa_{\myst{eff}}}{2\pi}& \int_{-\infty}^{u} du'\int_{S^2}d\Omega'  \, \frac{dE}{du'd\Omega'} \times
    \nonumber \\ 
    & \times \frac{\beta^2  \Lambda_{ij,kl}(\Omega) n'_k n'_l}{r(1-\beta \Vec{n}'\cdot \Vec{n})-v(u-u')\Vec{n}'\cdot \Vec{n}} \,.
\end{align}
We can now readily perform a limit to null infinity (defined as $r \rightarrow \infty$ while $u$ remains constant) to obtain the leading order expression for the gravitational wave memory: 
\begin{align}\label{eq:memorysolution}
    \delta h_{ij}^\text{TT} (x) & = \frac{\kappa_{\myst{eff}}}{2\pi r} \int du' d\Omega'  \frac{dE}{du'd\Omega'} \,  \frac{\beta^2  \Lambda_{ij,kl}(\Omega) n'_k n'_l}{1-\beta \Vec{n}'\cdot \Vec{n}(\Omega)} \,,\nonumber  
    \\  & = \frac{\kappa_{\myst{eff}}}{2\pi r}  \int d\Omega'\;\frac{dE(u)}{d\Omega'} \,  \left [ \frac{ \beta'_i \beta'_j}{1-\Vec{\beta}' \cdot \Vec{n}} \right ]^\text{TT} \,, 
\end{align}
where in the last step we have defined
\begin{equation}
    \frac{dE(u)}{d\Omega'}\equiv\int_{-\infty}^u du'\, \frac{dE}{du'd\Omega'}\,,
\end{equation}
as well as
\begin{equation}
    \beta'_i\equiv \beta n'_i=\beta n_i(\Omega')\,.
\end{equation}
We want to stress here, that due to the unique generic structure of the asymptotic energy-momentum tensor in Eq.~\eqref{eq:General Form asymptotic EMT}, the above limit to null infinity in fact does not depend on the precise form of the unbound energy flux, but is entirely governed by the Green's function.

\subsection{The Memory Effect in Horndeski Theory}
With the formula in Eq.~\eqref{eq:memorysolution} for the displacement memory associated to an arbitrary asymptotic energy momentum tensor, we can now assess the different types of memory arising in Horndeski theory by simply inserting the corresponding energy fluxes.

First of all, for the massless TT tensor modes traveling at the speed of light $\beta = 1$ the corresponding memory formula reads
\begin{align}
   \delta h_{ij}^{(h)\text{TT}}(u,r,\Omega)& = \frac{1}{8\pi r}  \int d\Omega' \; \frac{dE^{(h)}}{d\Omega'} \,  \left [ \frac{ n'_i n'_j}{1-\Vec{n}' \cdot \Vec{n}} \right ]^\text{TT} \, , \label{eq:metricmemory}
\end{align}
with
\begin{align}
   \frac{dE^{(h)}(u)}{d\Omega'} = \int_{-\infty}^u du' r^2\langle \partial_0 h_{kl}^\text{TT}\partial_0 h^{kl}_\text{TT}\rangle\,.
\end{align}
This follows from Eqs.~\eqref{eq:EMT h 00}, \eqref{eq:energy flux} and \eqref{eq:T00 h} of the unbound energy-momentum content as well as the value of the coefficient $k^{(h)}$ in Eq.~\eqref{eq:kappa h}. This expression agrees with the gravitational wave memory formula in GR \cite{Christodoulou:1991cr,PhysRevD.44.R2945,Favata:2010zu,Heisenberg:2023prj}.

Next, from Eq.~\eqref{eq:memorysolution} it is also possible to recover the well known formula for memory sourced by asymptotic matter fluxes, in particular of unbound single particle contributions of mass M and radial velocity $\beta$. In this case, the energy flux contribution is given by [Eq.~\eqref{eq:mattersourcesimp}]
\begin{align}
    \frac{dE^\text{(m)}}{du' d\Omega'}(x') = \beta r'^2 \langle\phantom{}_{\mys{(2)}}T_{00}\rangle = M\gamma \beta r'^2  \delta^{(3)}(\vec{x}'-\vec{x}''(t')) \,,
\end{align}
with $x''(t')$ denoting the trajectory of the particle. Moreover, for simplicity, we have dropped here the explicit averaging over small spacetime scales. However, this will miss an important property of the resulting memory, namely its characteristic rise-time that represents the natural high-frequency cutoff of the signal. Indeed, in the case of macroscopic classical matter, such a scale is naturally provided by the effective size of the distribution. On the other hand, the effective "size" of quantum particles related to their energies through the Heisenberg uncertainty principle would also in this case provide a natural rise-time scale of the resulting memory (see also \cite{maggiore2008gravitational,Allen:2019hnd}). In this work, we will however content ourselves with the point-like approximation. 

Since our goal is to study the memory effect in the limit $r \rightarrow \infty$, we only want to keep the radial component of the trajectory of the particle. By performing a change to spherical coordinates in which the delta function becomes
\begin{align}
    \delta^{(3)}(\Vec{x}' - \Vec{x}''(t')) = \frac{\delta (r' - r''(t'))}{r'^2} \delta^{(2)} (\Omega' - \Omega'') \,,
\end{align}
the trajectory is thus given by the radial motion $r''(t') = r''_0 + \beta t'$ at a fixed pair of angular coordinates $\Omega'' = \text{constant}$. In terms of the asymptotic retarded time $u'=t'-r'/\beta$, we thus have $r'' = r''_0 + \beta u' +r'$. Without loss of generality, we can set the constant $r''_0$ to be zero, such that the energy flux reads
\begin{align}
    \frac{dE^\text{(m)}}{du' d\Omega'}(x') &= M\gamma \beta  \delta (\beta u')\delta^{(2)}(\Omega' - \Omega'')  \nonumber \\
     & = M\gamma \delta (u') \frac{\delta (\theta' - \theta'')}{\sin \theta'} \delta (\phi' - \phi'') \,.
\end{align}
Plugging this expression into our general form of memory Eq.~\eqref{eq:memorysolution}, we obtain
\begin{align}
    \delta h^\text{(m)TT}_{ij} (x) = \frac{\kappa^\text{(m)}}{2 \pi r}  \frac{M}{\sqrt{1-v^2}}\Theta(u)\left [ \frac{\beta''_i \beta''_j}{1-\Vec{\beta}'' \cdot \Vec{n}}\right ]^\text{TT} \, , \label{eq:singlememory}
\end{align}
where here the source direction is given by the fixed angle $\Omega''$, $\Theta(u)$ is the Heaviside step function and
\begin{equation}
   \beta''_i=\beta n_i(\Omega'')\,.
\end{equation}
For a multi-particle system, the result is then just a sum of the memory contribution from each particle, 
\begin{align}
    \delta h^\text{(m)TT}_{ij} (x) = \frac{\kappa^\text{(m)}}{2 \pi r}\Theta(u)  \sum_{A=1}^{N} \frac{\pm M_A}{\sqrt{1-v_A^2}}\left [ \frac{\beta''_i{}^A \beta''_j{}^A}{1-\Vec{\beta}''_A \cdot \Vec{n}}\right ]^\text{TT} \,,\label{eq:linearmemory}
\end{align}
where incoming particles pick up an additional minus sign and it is understood that incoming particles are characterized by a negative radial velocity. Taking the difference between early and late times, we thus recover the known result from GR (see e.g. \cite{Braginsky:1987gvh,Thorne:1992sdb}) with [Eq.~\eqref{eq:kappa m}]
\begin{equation}
    \kappa^\text{(m)}=\kappa=8\pi G\,,
\end{equation}
since in this case $\bar G_4=1$.

Finally, the general formula in Eq.~\eqref{eq:memorysolution} also allows for the derivation of new results. In the case of Horndeski theory, for instance, through the memory formula associated to the massive scalar wave $\varphi$ emitted by the source. This asymptotic wave can be described as a superposition of radially outward plane waves of speed $\beta$ depending on the frequency content of the wave. Thus, each plane wave contributes to the memory through a contribution of the form
\begin{align}
    \delta h_{ij}^\text{($\varphi$)TT} (x) & = \frac{\kappa^\text{($\varphi$)}}{2\pi r}  \int d\Omega' \; \frac{dE^\text{($\varphi$)}}{d\Omega'} \,  \left [ \frac{ \beta'_i \beta'_j}{1-\Vec{\beta}' \cdot \Vec{n}} \right ]^\text{TT} \, , 
\end{align}
where $\kappa^\text{($\varphi$)}$ is given in Eq.~\eqref{eq:kappa phi} and [Eq.~\eqref{eq:T00 phi}]
\begin{align}
   \frac{dE^{(\varphi)}(u)}{d\Omega'} = \int_{-\infty}^u du' r^2\langle \partial_0 \varphi \partial_0 \varphi \rangle\,.
\end{align}

However, this only represents the contribution of the monochromatic plane wave modes with a given frequency, leading to the asymptotic radial velocity $\beta$ within the chosen source centered frame. Indeed, the group velocity of a massive field wave defined in Eq.~\eqref{eqn:phivelocity} depends on its frequency. Thus, just as for the multi-particle systems discussed above, through linearity, the total contribution is given by summing all the contributions from every plane wave frequency and the memory formula for a realistic emission of massive scalar waves is of the form
\begin{align}
    \delta h_{ij}^\text{($\varphi$)TT}  = \frac{\kappa^\text{($\varphi$)}}{4\pi r}  \int_0^1 d\beta  \int d\Omega' \; 
    \frac{dE^\text{($\varphi$)}(\beta)}{d\Omega'} \,  \left [ \frac{ \beta'_i \beta'_j}{1-\Vec{\beta}' \cdot \Vec{n}} \right ]^\text{TT} \, . \label{eq:phimemories}
\end{align}

The total displacement memory effect within Horndeski theory is given by the superposition of all the three contributions in Eqs.~\eqref{eq:metricmemory}, \eqref{eq:linearmemory} and \eqref{eq:phimemories}
\begin{align}
    \delta h_{ij}^\text{TT} (x) = \delta h_{ij}^\text{($h$)TT} (x) + \delta h_{ij}^\text{(m)TT} (x)+\delta h_{ij}^\text{($\varphi$)TT} (x)  \,.
\end{align}



\section{Discussion and Conclusion}\label{sec:Discussion}

Foremost, this paper provides a unifying framework to discuss ordinary and null memory within metric theories of gravity. With Horndeski theory as an explicit example, it is shown that the solution to the leading-order low-frequency equation within the Isaacson approach to gravitational waves only depends on a universal form of the coarse-grained source energy-momentum and thus provides a very general form of the contributions to the tensor displacement memory. 
Such an approach can also be applied to other scalar-tensor theories such as the DHOST theories \cite{Langlois_2019}, and the result is expected to hold the same form with different coefficients than Eq.~\eqref{eq:coeff}.

While the correspondence between the different types of memory were studied before for the specific example of the point particle approximation \cite{Thorne:1992sdb,Tolish:2014bka,Tolish:2014oda}, the existing individual derivations of the corresponding formulas for ordinary and null memory in the general case were rather distinct in nature. In contrast, the Isaacson approach not only unambiguously identifies memory contributions as low-frequency corrections to a primary energy-momentum emission, but provides the means of discussing all types of memory based on a single memory evolution equation, and in particular a continuous transition between null and ordinary memory.

The resulting general memory formula in Eq.~\eqref{eq:memorysolution} not only recovers known results but also includes the novel contribution given by the tensor memory sourced by the emission of a massive scalar wave in Horndeski gravity offered in Eq.~\eqref{eq:phimemories}. Massive scalar wave emissions represent a viable signature beyond GR as studied for instance in \cite{Silva:2017uqg,Silva:2020omi,Elley:2022ept,Degollado:2014vsa,Okawa:2014nda,Doneva:2019vuh,Richards:2023xsr}, such an extension of the memory formula to massive scalar fields represents an important step towards a complete study of beyond GR effects that might be captured by future memory measurements. However, while the memory signal from a nearly monochromatic scalar emission is expected to be unobservable, a direct application of the result in Eq.~\eqref{eq:phimemories} for a burst of massive scalar perturbations would first require a decomposition of the signal into a superposition of plane waves, each characterized by their own propagation speed. In this context, computing the size and shape of the memory contribution for a specific promising example would represent an interesting follow-up of this work.

Moreover, observe that a restriction to a constant asymptotic value of the additional scalar field in the gravitational sector ensures that the asymptotic Minkowski background preserves local Lorentz invariance. Another interesting extension of the present considerations would therefore be to analyze the consequences of giving up the requirement of asymptotic local Lorentz invariance.

In particular also vector-tensor theories with a non-trivial asymptotic value that (spontaneously) break local Lorentz invariance represent a natural description of such a situation. Thus, an extension of the present work to additional massive vector degrees of freedom within the gravitational sector of metric theories would be desirable. In this case, the gauge-invariant SVT decomposition introduced in Section \ref{Ssec:GaugeInv} might especially prove useful when decoupling the propagating modes of the theory.

Generally, it is expected that in such Lorentz breaking scenarios the luminality of the tensor waves is lost as well, therefore introducing an alternative causal structure on the asymptotic spacetime. In this context, the very broadly applicable Isaacson viewpoint of understanding asymptotic radiation and memory seems essential, as other methods relying on a well-defined boundary of the spacetime at null infinity would fail.

\begin{acknowledgments}
LH is supported by funding from the European Research Council (ERC) under the European Unions Horizon
2020 research and innovation programme grant agreement No 801781 and by the Swiss National Science Foundation
grant 179740. LH further acknowledges support from the Deutsche Forschungsgemeinschaft (DFG, German Research Foundation) under Germany's Excellence Strategy EXC 2181/1 - 390900948 (the Heidelberg STRUCTURES Excellence Cluster).

\end{acknowledgments}



\appendix

\section{Derivation of the Gauge Invariant Leading Order Propagation Equations}\label{Appendix:GaugeInvEq}
In this appendix, we derive the gauge-invariant linear propagation equations in Horndeski theory, as shown in Eqs.~\eqref{eqn:HorndeskiWaveEq} and \eqref{eqn:HorndeskiConstraintEq}. 


Consider first the leading order metric field equations at the high frequency level in Eq.~\eqref{eqn:highfreqgHorndeski}. Within the 3+1 and SVT decomposition in the asymptotic region presented in Sec.~\ref{Ssec:GaugeInv}, we can separately treat the different components of the equations of motion and rewrite them purely in terms of manifestly gauge invariant quantities. Firstly, the 00-component of the metric equation converts to
\begin{align}
    -\Bar{G}_{4} \nabla^2 \Theta   =\Bar{G}_{4,\phi} \nabla^2 \varphi \,.
\end{align}
On the other hand, the 0i-part of the metric equation can be further decomposed into its transverse and longitudinal components 
\begin{subequations}
\begin{align}
    -\frac{1}{2} \Bar{G}_{4} \nabla^2 \Xi^i &=0 \,, \\
    -\Bar{G}_{4} \partial^i \dot{\Theta}   &=\Bar{G}_{4,\phi} \partial^i \dot{\varphi} \,.
\end{align}
\end{subequations}
Finally, the spatial ij-component can be separated into its transverse-traceless, longitudinal, trace, and vector contributions:
\begin{subequations}
\begin{align}
    \frac{1}{2} \Bar{G}_{4} \Box E_{ij} &= 0 \,, \\
    -\Bar{G}_{4} \ddot{\Theta}   &=\Bar{G}_{4,\phi}  \ddot{\varphi} \,, \\
    \Bar{G}_{4} (-3 \ddot{\Theta} + \nabla^2 \Theta -2 \nabla^2 \Phi)  &=\Bar{G}_{4,\phi} (3 \ddot{\varphi} -2 \nabla^2 \varphi) \,, \\
    \Bar{G}_{4} (- \partial_{(i} \dot{\Xi}_{j)} ) & = 0 \,.
\end{align}
\end{subequations}
Here, the symbol $\nabla^2$ denotes the Laplacian operator.

Since the system is only physical when $\Bar{G}_4 \neq 0$, the above gauge invariant EOMs can be simplified by dividing out $\Bar{G}_4$. Employing the notation $\sigma \equiv \Bar{G}_{4,\phi} / \Bar{G}_4$, the independent components of the above field equations reduce to three constraint equations and one dynamical equation
\begin{subequations}
    \begin{align}
        \nabla^2 (\Theta +\sigma \varphi) = 0 \,, \\
        \nabla^2 \Xi^i = 0 \,, \\
        \nabla^2 (\Phi - \frac{1}{2}\sigma \varphi) = 0 \,, \\
        \Box E^\text{TT}_{ij} = 0 \,. 
    \end{align}
\end{subequations}

The last equation indicates that as in GR only the TT part of the high-frequency metric perturbation $h^H_{\mu\nu}\,$ is dynamical. 
The other three Laplacian equations unveil how the scalar field imprints on the metric. Because the scalar field and all gauge invariant quantities fulfill a well-behaved physical boundary condition at $r \rightarrow \infty$, the solutions of the Laplacian equations are simply given by 
\begin{subequations}
\begin{align}
    \Theta &= - \sigma \varphi \,, \\
    \Phi &= \frac{1}{2} \sigma \varphi, \\
    \Xi^i & = 0 \,.
\end{align}
\end{subequations}

Using the above solutions, we can in turn drastically simplify the leading order propagation equation of the scalar field perturbation in Eq.~\eqref{eqn:highfreqphiHorndeski} by replacing the gauge invariants $\Theta$ and $\Phi$ with scalar field perturbation $\varphi$. The scalar field propagating equation can then be written in terms of the gauge invariant quantities as
\begin{align}
     (\bar{G}_{2,X}-2\bar{G}_{3,\phi}+\frac{3(\bar{G}_{4,\phi})^2 }{\bar{G_4} }  ) (-\ddot{\varphi}+\nabla^2\varphi)+\varphi\bar{G}_{2,\phi\phi } =0 \,.
\end{align}
For the physically interesting solutions, we impose 
\begin{equation}
    \bar{G}_{2,X}-2\bar{G}_{3,\phi}+3(\bar{G}_{4,\phi})^2 / \bar{G_4}  \neq 0\,,
\end{equation}
and define the effective mass as
\begin{align}
    m^2 \equiv \frac{\bar{G}_{2,\phi\phi}}{\bar{G}_{2,X}-2\bar{G}_{3,\phi}+3(\bar{G}_{4,\phi})^2 / \bar{G_4} } \,. \label{eq:effmass}
\end{align}
The high-frequency scalar field equation then simplifies to the Klein-Gordon equation given in Eq.~\eqref{eqn:HorndeskiWaveEq}.

